\documentclass[submission,copyright,creativecommons]{eptcs}
\providecommand{\event}{NCMA 2024} 
\usepackage[strings]{underscore}   

\usepackage{amssymb}
\usepackage{amsmath}
\usepackage{shuffle}
\usepackage{amsfonts}
\usepackage[all]{xy}

\newcommand{\Accept}{{\sf Accept}}
\newcommand{\Reject}{{\sf Reject}}

\newcommand{\beginproof}{{\noindent \bf Proof.~}}
\newcommand{\myendproof}{\hspace*{\fill} $\Box$ \vspace{+0.2cm}}

\newtheorem{lemma}{Lemma}{\bf}{\it}
\newtheorem{proposition}[lemma]{Proposition}{\bf}{\it}
\newtheorem{theorem}[lemma]{Theorem}{\bf}{\it}
\newtheorem{corollary}[lemma]{Corollary}{\bf}{\it}
\newtheorem{definition}[lemma]{Definition}{\bf}{\it}
\newtheorem{example}[lemma]{Example}{\bf}{\rm}
{\bf}{\rm}
{\bf}{\it}

\title{Repetitive Finite Automata With Translucent Letters}
\author{Franti{\v s}ek Mr\'az
\institute{Charles University\\
Department of Computer Science\\
Malostransk\'e n\'am.~25\\
118 00 PRAHA, Czech Republic}
\email{frantisek.mraz@mff.cuni.cz}
\and
Friedrich Otto
\institute{Universit\"at Kassel\\
Fachbereich Elektrotechnik/Informatik\\
34109 KASSEL, Germany}
\email{\quad f.otto@uni-kassel.de}
}

\def\titlerunning{Repetitive Finite Automata with Translucent Letters}
\def\authorrunning{F.~Mr\'az and F.~Otto}

\begin{document}

\maketitle

\begin{abstract}
Here we propose an extension of the (deterministic and the nondeterministic) finite automaton with translucent letters (DFAwtl and NFAwtl),
which lies between these automata and their non-returning variants (that is, the nr-DFAwtl and the nr-NFAwtl).
This new model works like a DFAwtl or an NFAwtl,
but on seeing the end-of-tape marker, it
may change its internal state and continue with its computation instead of just ending it, accepting or rejecting.
This new type of automaton is called a \emph{repetitive deterministic} or \emph{nondeterministic finite automaton with translucent letters}
(\emph{RDFAwtl} or \emph{RNFAwtl}).
In the deterministic case, the new model is strictly more expressive than the DFAwtl,
but less expressive than the nr-DFAwtl,
while in the nondeterministic case, the new model is equivalent to the NFAwtl.
\end{abstract}

\section{Introduction}\label{sec0}
While a finite automaton reads its input strictly from left to right, letter by letter,
by now many types of automata have been considered in the literature that process their inputs in a different, more involved way.
Under this aspect, the most extreme is the \emph{jumping finite automaton} of Meduna and Zemek~\cite{MeZe2012} (see also~\cite{FPS2015}),
which, after reading a letter, jumps to an arbitrary position of the remaining input.
It is known that the jumping finite automaton accepts languages that are not even context-free,
like the language $\{\,w\in\{a,b,c\}^*\mid |w|_a = |w|_b = |w|_c\,\}$,
but at the same time, it does not even accept the finite language~$\{ab\}$.

Another example is the \emph{restarting automaton} as introduced by Jan{{\v c}}ar, Mr{\'a}z, Pl{\'a}tek, and Vogel in~\cite{JMPV95},
which processes a given input in cycles.
In each cycle, a restarting automaton scans its tape contents from left to right,
using a window of a fixed finite size, until it executes a delete/restart operation.
Such an operation deletes one or more letters from the current contents of the window,
returns the window to the left end of the tape,
and resets the automaton to its initial state.
If a window of size larger than one is used, these so-called R-automata accept a proper superclass of the regular languages
that is incomparable to the context-free and the growing context-sensitive languages 
with respect to inclusion (see, e.g.,~\cite{otto139}).
However, with a window of size one, the R-automata accept exactly the regular languages~\cite{Mra01}.

Finally, there is the (deterministic and nondeterministic) finite automaton \emph{with translucent letters} (or DFAwtl and NFAwtl) of Nagy and Otto~\cite{otto185},
which is equivalent to a cooperating distributed system of stateless deterministic R-automata with windows of size one.
For each state $q$ of an NFAwtl, there is a set $\tau(q)$ of \emph{translucent letters},
which is a subset of the input alphabet that contains those letters that the automaton cannot see when it is in state~$q$.
Accordingly, in each step, the NFAwtl just reads (and deletes) the first letter from the left which it can see,
that is, which is not translucent for the current state.
Here, it is important to notice that, in each step, an NFAwtl reads the current tape contents from the very left,
that is, after deleting a letter, it returns its head to the first letter of the remaining tape contents.
It has been shown that the NFAwtl accepts a class of semi-linear languages
that properly contains all rational trace languages,
while its deterministic variant, the DFAwtl,
is properly less expressive.
In fact, the DFAwtl just accepts a class of languages that is incomparable to the rational trace languages with respect to inclusion~\cite{otto176,NaOtLATA2011,otto195,otto206}.
Although the NFAwtl is quite expressive, it cannot even accept the deterministic linear language
$L_{2} = \{\,a^nb^n\mid n\ge 0\,\}$, as such an automaton cannot possibly compare the number of occurrences
of the letter $a$ with the number of occurrences of the letter $b$ and ensure, at the same time, that all $a$'s
precede the first~$b$.

To make up for this shortcoming,
a variant of the finite automaton with translucent letters has been proposed in~\cite{otto268},
which, after reading and deleting a letter, does not return its head to the first letter of the remaining tape contents,
but that rather continues from the position of the letter just deleted.
This means that, in general, only a scattered subword of the input has been read and deleted
before the head reaches the end of the input.
If the computation is now required to halt, either accepting or rejecting,
then it can easily be shown that this type of automaton just accepts the class of regular languages.
For the right one-way jumping finite automaton of~\cite{BiHo_IC284,CFY_IJFCS27},
this problem is overcome by cyclically shifting all the translucent letters that are encountered during a computation
to the end of the current tape contents.
In this way, these letters may be read and deleted at a later stage of the computation.
For the type of automaton proposed in~\cite{otto268}, a different approach was taken.
When the head of the automaton reaches the end of the input,
which is marked by a special end-of-tape marker, then the automaton can decide whether to accept, reject, or continue,
which means that it changes its state and again reads the remaining tape contents from the beginning.

It has been established that this type of automaton, called a \emph{non-returning finite automaton with translucent letters} or an \emph{nr-NFAwtl},
is strictly more expressive than the NFAwtl.
This result also holds for the deterministic case, although
the deterministic variant, the \emph{nr-DFAwtl},
is still not sufficiently expressive to accept all rational trace languages.
In~\cite{otto270}, the nr-DFAwtl and the nr-NFAwtl are compared to the jumping finite automaton,
the right one-way jumping finite automaton of~\cite{BiHo_IC284,CFY_IJFCS27}, and the
right-revolving finite automaton of~\cite{BBHK_IC207}, deriving the complete taxonomy of the resulting
classes of languages.
As it turns out, the nr-DFAwtl can be seen as an extension of the
right one-way jumping finite automaton that can detect the end of its input.

When we look at the above description of the generalization of the NFAwtl to the nr-NFAwtl,
then we realize that this generalization actually consists of two steps:
\begin{itemize}
 \item The head of the automaton does not return to the left end of the tape after a letter has been read and deleted.
 This is the \emph{non-returning} property (in contrast to the \emph{returning} property of the NFAwtl).
 \item Once the end-of-tape marker is reached, an nr-NFAwtl may execute a step that changes its state and returns
 its head to the left end of the tape.
 We call this the property of being \emph{repetitive} (in contrast to the \emph{non-repetitiveness}
 of the NFAwtl, which must immediately halt as soon as its head reaches the end-of-tape marker).
\end{itemize}

In this paper, we study the influence that these two properties have on the expressive capacity
of the finite automaton with translucent letters in detail.
As we consider both, the deterministic and non-deterministic variants of the resulting types of automata,
we obtain eight different classes of automata with translucent letters.
In addition, we also include the (deterministic and the non-deterministic)
right one-way jumping finite automaton in our study.
We shall derive the complete taxonomy of the resulting language classes,
which will nicely illustrate the effects that the non-returning property and the repetitiveness have.

Actually, we shall only encounter one new language class that has not been considered before:
the class $\mathcal{L}(\mbox{\sf RDFAwtl})$ of languages that are accepted by repetitive DFAwtls.
After presenting the necessary notation and definitions in Section~\ref{sec1},
we shall present our results on repetitive DFAwtls and repetitive NFAwtls in Section~\ref{sec2}.
Here, it turns out that the repetitive DFAwtl (the RDFAwtl) is strictly more expressive than the DFAwtl,
while its nondeterministic variant,
the repetitive NFAwtl (the RNFAwtl), is equivalent to the NFAwtl.
Then, in Section~\ref{sec3}, we consider closure and non-closure properties for the language
class $\mathcal{L}({\sf RDFAwtl})$.
Finally,
in the concluding section, we address the membership problem and some other decision problems
for the RDFAwtl in short, and we state some open problems.

\section{Definitions}\label{sec1}
First, we restate the definition of the nondeterministic finite automaton with translucent letters
and its deterministic variant, the DFAwtl, from~\cite{otto185}.

\begin{definition}\label{DefNFAwtl1}
A \emph{finite automaton with translucent letters}, or
an \emph{NFAwtl} for short,
is defined as a 7-tuple $A = (Q,\Sigma,\lhd,\tau,I,F,\delta)$,
where $Q$ is a finite set of internal states,
$\Sigma$ is a finite alphabet of input letters,
$\lhd\not\in\Sigma$ is a special letter that is used as an \emph{end-of-tape marker},
$\tau:Q\to \mathcal{P}(\Sigma)$ is a \emph{translucency mapping},\index{Translucency mapping}
$I\subseteq Q$ is a set of initial states,
$F\subseteq Q$ is a set of final states,
and $\delta:Q\times\Sigma\to \mathcal{P}(Q)$ is a \emph{transition relation}.
Here, we require that, for each state $q\in Q$ and each letter $a\in \Sigma$,
if $a\in\tau(q)$,
then $\delta(q,a)=\emptyset.$

An NFAwtl $A= (Q,\Sigma,\lhd,\tau,I,F,\delta)$ is a
\emph{deterministic finite automaton with translucent letters},
abbreviated as \emph{DFAwtl},\index{DFAwtl}
if $|I|=1$ and
$|\delta(q,a)|\le 1$ for all $q\in Q$ and all $a\in\Sigma$.
\end{definition}

A configuration of $A$ is a word from the set $Q\cdot\Sigma^*\cdot \lhd \,\cup\,\{\Accept,\Reject\}$.
A configuration of the form $qw\cdot\lhd$, where $q\in Q$ and $w\in\Sigma^*$,
expresses the situation that $A$ is in state~$q$,
its tape contains the word $w$ followed by the sentinel~$\lhd$,
and the head of $A$ is on the first letter of $w\cdot\lhd$.
For an input word $w\in\Sigma^*$,
a corresponding initial configuration is of the form $q_0w\cdot\lhd$, where $q_0\in I$.
The NFAwtl $A$ induces the following single-step computation relation on its set
of configurations:
$$qw\cdot\lhd \vdash_A \left\{ \begin{array}{ll}
     q'uv\cdot\lhd, & \mbox{if }w=uav,\,u\in(\tau(q))^*,\,a\in\Sigma\smallsetminus\tau(q),\,v\in\Sigma^*,\mbox{ and }q'\in\delta(q,a),\\
      \Reject,       & \mbox{if }w=uav,\, u\in(\tau(q))^*,\,
      a\in\Sigma\smallsetminus\tau(q),\,v\in\Sigma^*,\mbox{ and }\delta(q,a)=\emptyset,\\
     \Accept,      & \mbox{if }w\in(\tau(q))^* \mbox{ and }q\in F,\\
     \Reject,      & \mbox{if }w\in(\tau(q))^*\mbox{ and }q\not\in F.
     \end{array}\right.
$$
Thus, in each step, $A$ reads and deletes the first letter from the left that is not translucent for the current state.
In addition, if all letters on the tape are translucent for the current state, then $A$ halts, and it accepts
if the current state is final.
A word $w\in\Sigma^*$ is \emph{accepted by} $A$ if there exists an initial state $q_0\in I$
and a computation $q_0w\cdot \lhd \vdash_A^* \Accept$, where
$\vdash_A^*$ denotes the reflexive transitive closure of the single-step
computation relation~$\vdash_A$.
Now, $L(A) = \{\,w\in\Sigma^*\mid w\mbox{ is accepted by }A\,\}$
is the \emph{language accepted by}~$A$, and
$\mathcal{L}({\sf NFAwtl})$ denotes the class of all languages
that are accepted by~NFAwtls.
Analogously, $\mathcal{L}({\sf DFAwtl})$ denotes the class of all languages
that are accepted by DFAwtls.

Next, we define the first of the two possible extensions of the automaton with translucent letters
that we mentioned above.

\begin{definition}\label{DefRNFAwtl}
A \emph{repetitive} finite automaton with translucent letters, or an \emph{RNFAwtl},
is specified through a 6-tuple
$A=(Q,\Sigma,\lhd,\tau,I,\delta)$, where
$Q$, $\Sigma$, $\lhd$, $\tau$, and $I$ are defined as for an NFAwtl,
while
the transition relation $\delta$ is a mapping
$\delta:(Q\times (\Sigma\cup\{\lhd\})) \to \left(\mathcal{P}(Q)\cup\{\Accept\}\right)$.
Here, it is required that, for each state $q\in Q$ and each letter $a\in \Sigma$, $\delta(q,a)\subseteq Q$ and,
if $a\in\tau(q)$, then $\delta(q,a)=\emptyset.$
In addition, for each state $q\in Q$, $\delta(q,\lhd)$ is either a subset of $Q$ or $\delta(q,\lhd)=\Accept$.

The set of configurations for an RNFAwtl is the same as for an NFAwtl.
The RNFAwtl $A$ induces the following single-step computation relation on its set
of configurations:
$$qw\cdot\lhd \vdash_A \left\{\begin{array}{ll}
                               q'uv\cdot\lhd & \mbox{if } w=uav,\,u\in(\tau(q))^*, a\in\Sigma\smallsetminus\tau(q), v\in\Sigma^*,\mbox{ and } q'\in\delta(q,a),\\
                               \Reject        & \mbox{if } w=uav,\,u\in(\tau(q))^*, a\in\Sigma\smallsetminus\tau(q), v\in\Sigma^*,\mbox{ and } \delta(q,a)=\emptyset,\\
                               \Accept        & \mbox{if }w\in(\tau(q))^* \mbox{ and }\delta(q,\lhd) = \Accept,\\
                               \Reject        & \mbox{if }w\in(\tau(q))^* \mbox{ and }\delta(q,\lhd) = \emptyset,\\
                               q'w\cdot\lhd   & \mbox{if }w\in(\tau(q))^* \mbox{ and }q'\in\delta(q,\lhd).
                              \end{array}\right.$$
Thus, if all letters on the tape are translucent for the current state~$q$ and $\delta(q,\lhd)\subseteq Q$ is nonempty,
then~$A$ changes its state to $q'\in\delta(q,\lhd)$ and continues with its computation.
The language $L(A)$ accepted by $A$ is defined as
$L(A) = \{\,w\in\Sigma^*\mid w\mbox{ is accepted by }A\,\},$
that is, it consists of all words for which $A$ has an accepting computation, and
$\mathcal{L}({\sf RNFAwtl})$ denotes the class of all languages
that are accepted by~RNFAwtls.
\vspace{+2mm}

An RNFAwtl $A=(Q,\Sigma,\lhd,\tau,I,\delta)$ is a \emph{repetitive deterministic finite automaton with translucent letters},
or an \emph{RDFAwtl}, if $|I|=1$ and $|\delta(q,a)|\le 1$ for all $q\in Q$ and all $a\in\Sigma \cup\{\lhd\}$.
$\mathcal{L}({\sf RDFAwtl})$ denotes the class of all languages
that are accepted by RDFAwtls.
\end{definition}

Obviously, each NFAwtl can easily be turned into an RNFAwtl for the same language.
Of course, the same applies to a DFAwtl, giving an RDFAwtl for the same language.
Thus, we have the following inclusion relations.

\begin{proposition}\label{PropNFAinRNFA}
$\mathcal{L}({\sf DFAwtl}) \subseteq \mathcal{L}({\sf RDFAwtl})$ and
$\mathcal{L}({\sf NFAwtl}) \subseteq  \mathcal{L}({\sf RNFAwtl})$.
\end{proposition}

The following simple example illustrates
the way in which a repetitive finite automaton with translucent letters works. 

\begin{example}\label{ExLveec}
Let $A_{\vee,c} = (Q,\Sigma,\lhd,\tau,I,\delta)$ be the RDFAwtl that is defined as follows:
\begin{itemize}
 \item $Q=\{q_0,q_1,q_2,q_3,q_4,q_5,q_6,q_7\}$ and $I=\{q_0\}$,
 \item $\Sigma = \{a,b,c\}$,
 \item $\arraycolsep2pt
  \begin{array}[t]{lcllcllcllcl}
  \tau(q_0) & = & \{a,b\}, & \tau(q_1) & = & \emptyset, & \tau(q_2) & = & \{a\}, & \tau(q_3) & = & \{b\},\\
  \tau(q_4) & = & \emptyset, & \tau(q_5) & = & \{a\},& \tau(q_6) & = & \{a\},& \tau(q_7) & = & \{b\},
  \end{array}$
 \item $\arraycolsep=2pt
       \begin{array}[t]{lcllcllcllcl}
        \delta(q_0,c) & = & q_1, & \delta(q_0,\lhd) & = & q_4,\\
        \delta(q_1,a) & = & q_2, & \delta(q_1,b)    & = & q_3, & \delta(q_1,\lhd) & = & \multicolumn{4}{l}{\Accept,}\\
        \delta(q_2,b) & = & q_1, & \delta(q_3,a)    & = & q_1,\\
        \delta(q_4,a) & = & q_5, & \delta(q_4,b)    & = & q_7, & \delta(q_4,\lhd) & = & \multicolumn{4}{l}{\Accept,}\\
        \delta(q_5,b) & = & q_6, & \delta(q_6,b)    & = & q_4, & \delta(q_7,a) & = & q_6. 
       \end{array}$
       \newline
       \vspace{+1mm}

while $\delta$ yields the empty set for all other pairs from $Q\times(\Sigma\cup\{\lhd\})$.
\end{itemize}
Using the graphical notation introduced in~\cite{otto270} for describing non-returning NFAwtls,
the RDFAwtl~$A_{\vee,c}$ can be depicted more compactly by the diagram in Fig.~\ref{FigExLveec}.

\begin{figure}[ht]
\begin{center}
{\small
$$ \xymatrix @R=2pc@C=3pc{
      & *++[o][F-]{q_2}\ar@/^1pc/[r]^{(a^*,b)}
      & *++[o][F-]{q_1}\ar@/_1pc/[r]^{b} \ar@/^1pc/[l]_{a}\ar@/_1pc/[dr]^{\lhd}  & *++[o][F-]{q_3}\ar@/_1pc/[l]_{(b^*,a)} \\
  \ar[r]    &  *++[o][F-]{q_0}\ar@/_.5pc/[ru]_{(\{a,b\}^*,c)}\ar@/^.5pc/[rd]^{(\{a,b\}^*,\lhd)} & & *++[o][F-]{{\sf Accept}}\\
      & *++[o][F-]{q_5}\ar@/_1pc/[dr]_{(a^*,b)} & *++[o][F-]{q_4}\ar[r]^{b} \ar[l]_{a}\ar@/^1pc/[ur]^{\lhd} & *++[o][F-]{q_7}\ar@/^1pc/[dl]^{(b^*,a)}\\
      &  & *++[o][F-]{q_6}\ar[u]^{(a^*,b)} & &\\
} $$
}
\caption{The diagram describing the RDFAwtl $A_{\vee,c}$. An arrow from a state $q$ to a state $q'$ labeled with a single letter $x$ means that $\tau(q) = \emptyset$ and $q' \in \delta(q,x)$. An arrow labeled with a pair $(\Delta^*,x)$ means that $\tau(q)=\Delta$ and $q'\in \delta(q,x)$.}\label{FigExLveec}
\end{center}
\end{figure}
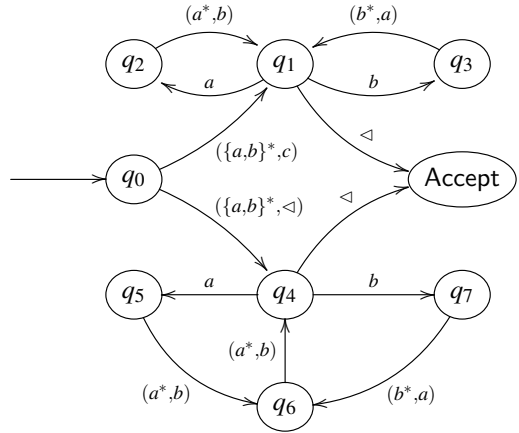

For example, the RDFAwtl $A_{\vee,c}$ can execute the following accepting computations:
$$
\arraycolsep=2pt
\begin{array}{lclclcl}
q_0aabbcba\cdot\lhd & \vdash_{A_{\vee,c}} & q_1aabbba\cdot\lhd & \vdash_{A_{\vee,c}} & q_2abbba\cdot\lhd & \vdash_{A_{\vee,c}} & q_1abba\cdot\lhd  \\
                   & \vdash_{A_{\vee,c}} & q_2bba\cdot\lhd & \vdash_{A_{\vee,c}} & q_1ba\cdot\lhd & \vdash_{A_{\vee,c}} & q_3a\cdot\lhd\\
                   & \vdash_{A_{\vee,c}} & q_1\cdot\lhd & \vdash_{A_{\vee,c}} & \Accept,
\end{array}$$
and
$$
\arraycolsep=2pt
\begin{array}{lclclcl}
q_0bbaabb\cdot\lhd & \vdash_{A_{\vee,c}} & q_4bbaabb\cdot\lhd & \vdash_{A_{\vee,c}} & q_7baabb\cdot\lhd & \vdash_{A_{\vee,c}} & q_6babb\cdot\lhd  \\
                   & \vdash_{A_{\vee,c}} & q_4abb\cdot\lhd & \vdash_{A_{\vee,c}} & q_5bb\cdot\lhd & \vdash_{A_{\vee,c}} & q_6b\cdot\lhd\\
                   & \vdash_{A_{\vee,c}} & q_4\cdot\lhd & \vdash_{A_{\vee,c}} & \Accept.
\end{array}$$
In fact, it can be checked that $$L(A_{\vee,c}) =
\{\,w\in\{a,b,c\}^*\mid |w|_c=1 \mbox{ and }|w|_a=|w|_b\,\}\,\cup \,\{\,w\in\{a,b\}^*\mid 2\cdot|w|_a=|w|_b\,\}.$$
\vspace{-12mm}

\hspace*{\fill}$\blacksquare$
\end{example}

Finally, we come to the second extension of the automaton with translucent letters
that we mentioned in the introduction.

\begin{definition}\label{DefNonret}
A \emph{non-repetitive non-returning} finite automaton with translucent letters,
or an \emph{nr-nr-NFAwtl},
is defined like an NFAwtl,
but its computation relation is defined differently.
Let $A = (Q,\Sigma,\lhd,\tau,I,F,\delta)$ be an nr-nr-NFAwtl.
Its set of configurations is $\Sigma^*\cdot Q\cdot\Sigma^*\cdot\lhd \,\cup\,\{\Accept,\Reject\}$.
A configuration of the form $xqw\cdot\lhd$, where $q\in Q$ and $x,w\in\Sigma^*$, expresses the situation that $A$ is in state~$q$,
the tape contains the word $xw\cdot\lhd$, and the head of $A$ is on the first letter of the suffix $w\cdot\lhd$.
The single-step computation relation that~$A$ induces on this set of configurations is defined as follows,
where $q\in Q$ and $x,w\in\Sigma^*$:
$$xqw\cdot\lhd \vdash_A \left\{ \begin{array}{ll}
     xuq'v\cdot\lhd, & \mbox{if }w=uav,\,u\in(\tau(q))^*,\,a\in\Sigma\smallsetminus\tau(q),\,v\in\Sigma^*,\mbox{ and }q'\in\delta(q,a),\\
      \Reject,       & \mbox{if }w=uav,\, u\in(\tau(q))^*,\,
      a\in\Sigma\smallsetminus\tau(q),\,v\in\Sigma^*,\mbox{ and }\delta(q,a)=\emptyset,\\
     \Accept,      & \mbox{if }w\in(\tau(q))^* \mbox{ and }q\in F,\\
     \Reject,      & \mbox{if }w\in(\tau(q))^*\mbox{ and }q\not\in F.
     \end{array}\right.
$$
Thus, in each step, $A$ reads and deletes the first letter to the right of the current position of its head that is not translucent for the current state.
In particular, after reading and deleting a letter, the head does \emph{not} return to the left end of the tape
(that is why this type of automaton is called \emph{non-returning}),
but it rather moves to the next letter.
In addition, if all letters on the tape that are to the right of the current head position are translucent for the current state,
then $A$ halts, and it accepts
if the current state is final.
A word $w\in\Sigma^*$ is \emph{accepted by} $A$ if there exists an initial state $q_0\in I$
and a computation $q_0w\cdot \lhd \vdash_A^* \Accept$, where
$\vdash_A^*$ denotes the reflexive transitive closure of the single-step
computation relation~$\vdash_A$.
Now, $L(A) = \{\,w\in\Sigma^*\mid w\mbox{ is accepted by }A\,\}$
is the \emph{language accepted by}~$A$, and
$\mathcal{L}(\mbox{\sf nr-nr-NFAwtl})$ denotes the class of all languages
that are accepted by~nr-nr-NFAwtls.

An nr-nr-NFAwtl $A=(Q,\Sigma,\lhd,\tau,I,F,\delta)$ is a
\emph{non-repetitive non-returning deterministic finite automaton with translucent letters},
or an \emph{nr-nr-DFAwtl}, if $|I|=1$ and $|\delta(q,a)|\le 1$ for all $q\in Q$ and all $a\in\Sigma$.
Then, $\mathcal{L}(\mbox{\sf nr-nr-DFAwtl})$ denotes the class of all languages
that are accepted by nr-nr-DFAwtls.
\end{definition}

The following result states that the non-repetitive non-returning finite automata with translucent letters of Def.~\ref{DefNonret}
are very weak in that they just accept the regular languages.

\begin{theorem}\label{ThmNrNFAwtl}
From an nr-nr-NFAwtl $A$, one can construct an NFA $B$ such that $L(B)=L(A)$.
In addition, if $A$ is deterministic, then so is $B$.
\end{theorem}

\beginproof
Let $A=(Q,\Sigma,\lhd,\tau,I,F,\delta)$ be an nr-nr-NFAwtl.
From the definition of the computation relation of~$A$, we see immediately that,
whenever $quav\cdot\lhd \vdash_A uq'v\cdot\lhd$ is a step in an accepting computation of $A$,
where $q,q'\in Q$, $u\in(\tau(q))^+$, and $a\in\Sigma$,
then the prefix $u$ will not be read again during the remaining part of this accepting computation.
Thus, instead of ignoring these letters, we could simply delete them.
Accordingly, $A$ accepts the same language as the NFA $B=(Q,\Sigma,I,F,\delta_B)$
that is defined through the following transition relation:
$$\delta_B(q,a) = \left\{ \begin{array}{ll}
                          q, & \mbox{if }a\in\tau(q),\\
                          \delta(q,a), &\mbox{if }a\not\in\tau(q).
                         \end{array}\right.$$
Trivially, if $A$ is deterministic, the constructed automaton $B$ is deterministic, too.
\myendproof

It was actually this observation that led us to define the nr-NFAwtl and the nr-DFAwtl in~\cite{otto268}.

\begin{definition}[\cite{otto268}]\label{DefnrDFAwtl}
An \emph{nr-NFAwtl} $A=(Q,\Sigma,\lhd,\tau,I,\delta)$ is defined like an RNFAwtl
but with the additional extension that it is non-returning,
that is, it behaves like an nr-nr-NFAwtl, but for some states $q\in Q$, $\delta(q,\lhd)$ may be a subset of~$Q$.
Thus, if $A$ is in a configuration of the form $xqw\cdot\lhd$, where $q\in Q$ satisfying $\delta(q,\lhd)\subseteq Q$,
$x\in\Sigma^*$, and $w\in (\tau(q))^*$,
then $xqw\cdot\lhd \vdash_A q'xw\cdot\lhd$ for each state $q'\in \delta(q,\lhd)$.
If $|I|=1$ and $|\delta(q,a)|\le 1 $ for all $q\in Q$ and all $a\in\Sigma\cup\{\lhd\}$,
then $A$ is an \emph{nr-DFAwtl}.
We use  $\mathcal{L}(\mbox{\sf nr-NFAwtl})$ and $\mathcal{L}(\mbox{\sf nr-DFAwtl})$
to denote the corresponding classes of languages.
\end{definition}

Thus, an nr-NFAwtl is both at the same time, non-returning in the sense of Def.~\ref{DefNonret} and repetitive in the sense of Def.~\ref{DefRNFAwtl}.
In particular, the nr-NFAwtl (nr-DFAwtl) should not be confused with the nr-nr-NFAwtl (nr-nr-DFAwtl) of Def.~\ref{DefNonret}.
The latter has been introduced here only to complete the picture.
Theorem~\ref{ThmNrNFAwtl} above shows that they are not really interesting types of automata with translucent letters.
Finally, we should also mention another related type of automaton,
the right one-way jumping finite automaton.

\begin{definition}[\cite{BiHo_IC284,CFY_IJFCS27}]\label{DefNROWJFA}
A \emph{nondeterministic right one-way jumping finite automaton}, or an \emph{NROWJFA},
is given through a 6-tuple $J = (Q,\Sigma,\lhd,I,F,\delta)$,
where $Q$, $\Sigma$, $\lhd$, $I$, and $F$ are defined as for an NFAwtl,
and $\delta:Q\times \Sigma \to \mathcal{P}(Q)$ is a transition relation.
For each state $q\in Q$, let $\Sigma_q = \{\,a\in\Sigma\mid \delta(q,a) \not= \emptyset\,\}$
be the set of letters that $J$ can read in state~$q$.

A configuration of the NROWJFA $J$ is a word $qw\cdot\lhd$ from the set $Q\cdot\Sigma^*\cdot\lhd$.
The \emph{computation relation}~$\circlearrowright^*_J$ that~$J$ induces on its set of configurations
is the reflexive and transitive closure of the \emph{right one-way jumping relation} $\circlearrowright_J$
that is defined as follows, where $q,q'\in Q$, $x,y\in\Sigma^*$, and $a\in\Sigma:$
$$qxay\cdot\lhd \circlearrowright_J q'yx\cdot\lhd \mbox{ if }x\in (\Sigma\smallsetminus \Sigma_q)^* \mbox{ and }q'\in\delta(q,a).$$
Thus, being in state $q$, $J$ reads and deletes the first letter to the right of the actual head position
that it can actually read in that state, while the prefix that consists of letters for which
$J$ has no transitions in the current state is cyclically shifted
to the end of the current tape inscription.
Then,
$$L(J) = \{\,w\in\Sigma^*\mid \exists\, q_0\in I\,\exists\, q_f\in F: q_0w\cdot\lhd \circlearrowright^*_J q_f\cdot\lhd\,\}$$
is the language accepted by the NROWJFA~$J$.
\vspace{+2mm}

The NROWJFA $J$ is \emph{deterministic}, that is, a \emph{right one-way jumping finite automaton} or an \emph{ROWJFA},
if $|I|=1$ and $|\delta(q,a)|\le 1$ for all $q\in Q$ and $a\in \Sigma$.
\end{definition}

Actually, as defined in~\cite{BiHo_IC284,CFY_IJFCS27}, the (N)ROWJFA does not have an end-of-tape marker,
but it is obvious that our definition is equivalent to the original one.
We just introduced this end-of-tape marker to ensure consistency with our other types of automata.
We see that the (N)ROWJFA overcomes the problem of processing letters that are skipped over
by cyclically shifting these letters to the right so that they can be read later.
The following results are known concerning the various types of automata introduced above.

\begin{theorem}[\cite{otto270}]\label{ThmnrNFA}
$\begin{array}[t]{clcccccc}
{\rm (a)} & {\sf REG} & \subsetneq & \mathcal{L}(\mbox{\sf DFAwtl}) & \subsetneq & \mathcal{L}(\mbox{\sf nr-DFAwtl}) & \subsetneq & \mathcal{L}(\mbox{\sf nr-NFAwtl}).\\
{\rm (b)} & {\sf REG} & \subsetneq & \mathcal{L}(\mbox{\sf DFAwtl}) & \subsetneq & \mathcal{L}(\mbox{\sf NFAwtl}) & \subsetneq & \mathcal{L}(\mbox{\sf nr-NFAwtl}).\\
{\rm (c)} &           &            & \mathcal{L}(\mbox{\sf ROWJFA}) & \subsetneq & \mathcal{L}(\mbox{\sf nr-DFAwtl}).\\
{\rm (d)} &           &            & \mathcal{L}(\mbox{\sf NROWJFA}) & \subsetneq & \mathcal{L}(\mbox{\sf nr-NFAwtl}).\\
{\rm (e)} &           &            & \mathcal{L}(\mbox{\sf ROWJFA}) & \subsetneq & \mathcal{L}(\mbox{\sf NROWJFA}).\\
\end{array}$
\end{theorem}

In addition, $\mathcal{L}(\mbox{\sf ROWJFA})$ is incomparable under inclusion to $\mathcal{L}(\mbox{\sf DFAwtl})$ and $\mathcal{L}(\mbox{\sf NFAwtl})$,
and $\mathcal{L}(\mbox{\sf NROWJFA})$ is incomparable to $\mathcal{L}(\mbox{\sf DFAwtl})$, $\mathcal{L}(\mbox{\sf NFAwtl})$,
and $\mathcal{L}(\mbox{\sf nr-DFAwtl})$.
Thus, it remains to compare the deterministic and nondeterministic repetitive finite automata with translucent letters
to these other types of automata.

\section{Comparing the Repetitive Automata to the Non-Repetitive Ones}\label{sec2}

We claim that the language $L_{\vee,c}=L(A_{\vee,c})$ of Example~\ref{ExLveec} is not accepted by any DFAwtl.
As each DFAwtl can be simulated by an RDFAwtl,
this shows that $\mathcal{L}({\sf DFAwtl}) \subsetneq \mathcal{L}({\sf RDFAwtl })$.

\begin{lemma}\label{LemLveec}
$L_{\vee,c}\not\in\mathcal{L}({\sf DFAwtl})$.
\end{lemma}

\noindent
{\bf Proof.}
Assume that $A=(Q,\Sigma,\lhd,\tau,I,F,\delta)$ is a DFAwtl
that accepts the language~$L_{\vee,c}$,
where $Q=\{q_0,q_1,\ldots,q_{m-1}\}$, $\Sigma=\{a,b,c\}$, and $I=\{q_0\}$.

Let $n>2m$, and let $w=a^nb^nc\in L_{\vee,c}$.
Then the computation of $A$ on input $w$ is accepting,
that is, it is of the form
$$q_0a^nb^nc\cdot\lhd \vdash_{A} q_{i_1}w_1\cdot\lhd\vdash_{A}\cdots\vdash_{A} q_{i_r}w_r\cdot\lhd \vdash_{A}\Accept,$$
where $w_r\in(\tau(q_{i_r}))^*$ and $q_{i_r}\in F$.
If $|w_r|_a > 0$, then $A$ would also accept on input $a^{n+1}b^nc\not\in L_{\vee,c}$,
if $|w_r|_b > 0$, then $A$ would also accept on input $a^nb^{n+1}c\not\in L_{\vee,c}$,
and if $|w_r|_c > 0$, then $A$ would also accept on input $a^nb^n\not\in L_{\vee,c}$.
Hence, it follows that $w_r=\lambda$, that is, the accepting computation above consists of $2n+1$ transition steps,
each of which deletes a letter, and the final accepting step.
In particular, the only occurrence of the letter $c$ is read and deleted during the above computation, that is,
there exist an index $j$ and integers $r,s\ge 0$ such that $r+s=j$ and
$$q_0a^nb^nc\cdot\lhd \vdash_A^{j} q_{i_j}a^{n-r}b^{n-s}c\cdot\lhd \vdash_A q_{i_{j+1}}a^{n-r}b^{n-s}\cdot\lhd
\vdash_A^* q_{i_r}\cdot\lhd \vdash_A \Accept.$$

We now distinguish several cases.
\begin{enumerate}
 \item[(1)]
Assume that $a,b\in\tau(q_{i_j})$.
Then $A$ executes the following computation on input $a^nb^{2n}$:
$$q_0a^nb^{2n}\cdot\lhd \vdash_A^j
q_{i_j}a^{n-r}b^{2n-s}\cdot\lhd \vdash_A \left\{\begin{array}{ll}
                                          \Accept, & \mbox{if }q_{i_j}\in F,\\
                                          \Reject, & \mbox{if }q_{i_j}\not\in F.
                                          \end{array}\right.$$
As $a^nb^{2n}\in L_{\vee,c}$, we see that the latter computation must be accepting, that is, $q_{i_j}\in F$.
Thus, $A$ can also execute the following accepting computation:
$$q_0a^{r+s+1}b^{r+s+1}\cdot\lhd \vdash_A^j q_{i_j}a^{s+1}b^{r+1}\cdot\lhd \vdash_A\Accept,$$
which, however, contradicts the fact that $a^{r+s+1}b^{r+s+1}\not\in L_{\vee,c}$.
\item[(2)] Assume that $a\not\in\tau(q_{i_j})$, but $b\in \tau(q_{i_j})$.
Then, $r=n$ and $s\le n$, that is, $a^{n-r}b^{n-s}c = b^{n-s}c$.
Now, $A$ executes the following computation on input $a^nb^{2n}$:
$$q_0a^nb^{2n}\cdot\lhd \vdash_A^j q_{i_j}a^{n-r}b^{2n-s}\cdot\lhd =
q_{i_j}b^{2n-s}\cdot\lhd \vdash_A \left\{\begin{array}{ll}
                                          \Accept, & \mbox{if }q_{i_j}\in F,\\
                                          \Reject, & \mbox{if }q_{i_j}\not\in F.
                                          \end{array}\right.$$
As $a^nb^{2n}\in L_{\vee,c}$, we see that the latter computation must be accepting, that is, $q_{i_j}\in F$.
Thus, $A$ can also execute the following accepting computation:
$$q_0a^{n}b^{3n+s}\cdot\lhd \vdash_A^j q_{i_j}b^{3n}\cdot\lhd \vdash_A\Accept,$$
which, however, contradicts the fact that $a^{n}b^{3n+s}\not\in L_{\vee,c}$.
\item[(3)] Assume that $b\not\in \tau(q_{i_j})$.
Then $r\le n$ and $s=n$, that is, $a^{n-r}b^{n-s}c = a^{n-r}c$.
As $n>m$, there exist a state $q$ and integers $k_0,k_1,t_0\ge 0$ and $t_1\ge 1$
such that the accepting computation above has the form
$$q_0a^nb^nc\cdot\lhd \vdash_A^* qa^{n-k_0}b^{n-t_0}c\cdot\lhd \vdash_A^+ qa^{n-k_0-k_1}b^{n-t_0-t_1}c\cdot\lhd
\vdash_A^* q_{i_j}a^{n-r}c\cdot\lhd\vdash_A^*\Accept.$$
Hence, we also obtain the following accepting computation:
$$q_0a^{n+k_1}b^{n+t_1}c\cdot\lhd \vdash_A^* qa^{n+k_1-k_0}b^{n+t_1-t_0}c\cdot\lhd \vdash_A^+ qa^{n-k_0}b^{n-t_0}c\cdot\lhd
\vdash_A^*\Accept.$$
This implies that $a^{n+k_1}b^{n+t_1}c\in L_{\vee,c}$, which yields $k_1=t_1$.

Now we consider the computations of $A$ on input $a^{n+\nu\cdot t_1}b^{n+\nu\cdot t_1}$
for all $\nu\ge 0$:
$$q_0a^{n+\nu\cdot t_1}b^{n+\nu\cdot t_1}\cdot\lhd
\vdash_A^* q_{i_j}a^{n-r}\cdot\lhd.$$
As $a^{n+\nu\cdot t_1}b^{2n+2\nu\cdot t_1}\in L_{\vee,c}$,
we see that the computation of $A$ that begins with the configuration
$q_{i_j}a^{n-r}b^{n+\nu\cdot t_1}$
leads to acceptance for all $\nu\ge 0$.
Hence, we obtain
$$q_0a^{n}b^nb^{n+ t_1}\cdot\lhd \vdash_A^* q_{i_j}a^{n-r}b^{n+t_1}\cdot\lhd \vdash_A^* \Accept,$$
but we have $a^nb^{n}b^{n+t_1}\not\in L_{\vee,c}$, as $t_1 >0$, a contradiction.
\end{enumerate}

As this covers all cases, we see that the language $L_{\vee,c}$ is indeed not accepted by any DFAwtl.
\hspace*{\fill}$\Box$
\vspace{+2mm}

On the other hand, it can be shown quite easily
that the RDFAwtl (the RNFAwtl) is a special case of the nr-DFAwtl (the nr-NFAwtl).

\begin{lemma}\label{LemRDFAinnrDFA}
From an RNFAwtl $A$, one can construct an nr-NFAwtl $B$ such that $L(B)=L(A)$.
In addition, if $A$ is deterministic, then so is $B$.
\end{lemma}

\beginproof
Let $A = (Q,\Sigma,\lhd,\tau,I,\delta)$ be an RNFAwtl.
We define an nr-NFAwtl $B = (Q_B,\Sigma,\lhd,\tau_B,I_B,\delta_B)$ that simulates the computations of $A$
as follows:
\begin{itemize}
 \item $Q_B = Q \cup \{\,q'\mid q\in Q\,\}$, where for each state $q\in Q$, $q'$ is an additional auxiliary state, and $I_B=I$,
 \item for each state $q\in Q$, $\tau_B(q) = \tau(q)$ and $\tau_B(q')=\Sigma$,
 \item for each state $q\in Q$ and each letter $a\in \Sigma$,
 $\delta_B(q,a)= \{\,p'\mid p\in \delta(q,a)\,\}$ and $\delta_B(q',a)=\emptyset$.
 \item
 Furthermore, for each state $q\in Q$, $\delta_B(q,\lhd) = \Accept$, if $\delta(q,\lhd) = \Accept$,
 and $\delta_B(q,\lhd) = \emptyset$, otherwise.
 Finally, $\delta_B(q',\lhd) = \{q\}$.
\end{itemize}
It remains to verify that $B$ just simulates the computations of~$A$.

Assume that $qw\cdot\lhd$ is a configuration of~$A$, that is, $q\in Q$ and $w\in\Sigma^*$.
From the definition of the computation relation $\vdash_A$, we see that there are four different cases that we must consider.

First, if $w=uav$ for some words $u\in(\tau(q))^*$, $v\in\Sigma^*$, and a letter $a\in(\Sigma\smallsetminus\tau(q))$,
then $A$ executes a transition from $\delta(q,a)$.
\begin{itemize}
\item If $p\in\delta(q,a)$, then $qw\cdot\lhd = quav\cdot\lhd\vdash_A puv\cdot\lhd$ is a possible step of~$A$.
In this case, $B$ can execute the following sequence of steps:
$$qw\cdot\lhd = quav\cdot\lhd \vdash_B up'v\cdot\lhd \vdash_B puv\cdot\lhd.$$
\item On the other hand, if $\delta(q,a)=\emptyset$, then $A$ halts and rejects.
However, in this case, also $\delta_B(q,a)=\emptyset$, and hence, $B$ halts and rejects as well.
\end{itemize}

Finally, if $w\in(\tau(q))^*$,
then $A$ halts.
\begin{itemize}
\item If $\delta(q,\lhd) = \Accept$, then $\delta_B(q,\lhd)=\Accept$, too.
\item If $\delta(q,\lhd) = \emptyset$, then $A$ rejects. In this case, $\delta_B(q,\lhd) = \emptyset$,
that is, $B$ halts and rejects as well.
\end{itemize}

It follows that $L(A) \subseteq L(B)$.
\vspace{+2mm}

Conversely, if $w\in L(B)$, then it is easily verified
that each accepting computation of $B$ on input $w$ is just a simulation
of an accepting computation of $A$ on input~$w$.
Thus, we see that $L(B)=L(A)$.

Finally, the above definition of $B$ shows that $B$ is deterministic, if $A$ is.
\myendproof

The language
$$L_\vee = \{\,w\in\{a,b\}^* \mid |w|_b = |w|_a \mbox{ or }|w|_b = 2\cdot|w|_a\,\}$$
is a rational trace language, and as such, it is accepted by an NFAwtl.
However, as proved in~\cite{otto270},
this language is not accepted by any nr-DFAwtl.
Hence, $L_\vee$ is not accepted by any RDFAwtl, either.
Thus, we immediately obtain the following non-inclusion result.

\begin{corollary}\label{CorLvee}
$\mathcal{L}({\sf NFAwtl}) \not\subseteq \mathcal{L}({\sf RDFAwtl})$.
\end{corollary}

As defined above, an RNFAwtl $A=(Q,\Sigma,\lhd,\tau,I,\delta)$ may run into an infinite computation.
Just assume that $q$ is a state of~$A$, $w\in (\tau(q))^*$, and $q\in \delta(q,\lhd)$.
Then $qw\cdot\lhd \vdash_A qw\cdot\lhd \vdash_A qw\cdot\lhd$, and so forth.
However, we can avoid this by converting $A$ into an equivalent RNFAwtl $B$ as follows.
\vspace{+2mm}

Let $B=(Q',\Sigma,\lhd,\tau',I',\delta')$, where $Q' = \{\,(q,S) \mid q\in Q\mbox{ and }S\subseteq Q\,\}$,
$I'=\{\,(q,\emptyset)\mid q\in I\, \}$,
$\tau'(q,S) = \tau(q)$ for all $q\in Q$ and all $S\subseteq Q$,
$$\delta'((q,S),a) = \{\,(p,\emptyset) \mid p\in\delta(q,a)\,\} \text{ for all $q\in Q$, $S\subseteq Q$, and all $a\in\Sigma$,}$$
and
$$\delta'((q,S),\lhd)  = \{\,(p,S\cup\{q\})\mid p\in\delta(q,\lhd)\mbox{ and } q\not\in S\,\}
\text{ for all $q\in Q$ and all $S\subseteq Q$}.$$
Finally, take $\delta'((q,S),\lhd) = \Accept$ if $\delta(q,\lhd) = \Accept$.
The set $S$ is used to record those states in which the end-of-tape marker has been reached,
and the computation has continued.
In the next cycle, when a non-translucent letter is read, then this set is emptied, otherwise,
the next state is added to it. This process continues until either a letter is read and deleted, or until
no new state can be added to the current set $S$, in which case the computation fails.
\vspace{+2mm}

Moreover, an RNFAwtl $A=(Q,\Sigma,\lhd,\tau,I,\delta)$ may accept without having read and deleted its input completely.
However, we can easily extend the RNFAwtl $A$ into an equivalent RNFAwtl $C$ that always reads and deletes its input completely before it accepts.
Just take $C=(Q\cup\{q_e\},\Sigma,\lhd,\tau',I,\delta')$, where $q_e$ is a new state,
$\tau'(q)=\tau(q)$ for all $q\in Q$ and $\tau'(q_e)=\emptyset$, and $\delta'$ is defined as follows:
$$\begin{array}{clcll}
- & \delta'(q,a) & = & \delta(q,a) & \mbox{for all }q\in Q\mbox{ and all }a\in\Sigma,\\
- & \delta'(q,\lhd) & = & \multicolumn{2}{l}{\left\{\begin{array}{ll}
                                 \delta(q,\lhd), & \mbox{if }\delta(q,\lhd)\not=\Accept,\\
                                 \{q_e\},        & \mbox{if }\delta(q,\lhd) = \Accept,
                                 \end{array}\right.}\\
- & \delta'(q_e,a) & = & \{q_e\} & \mbox{for all }a\in\Sigma,\\
- & \delta'(q_e,\lhd) & = & \Accept.
\end{array}$$
Given a word $w\in\Sigma^*$ as input, the RNFAwtl $C$ will execute exactly the same steps
as the RNFAwtl~$A$ until~$A$ accepts.
Now, the accept step of~$A$ is simulated by $C$ through changing into state~$q_e$.
As $\tau'(q_e) = \emptyset$ and as $\delta'(q_e,a)=\{q_e\}$ for all $a\in\Sigma$,
$C$ will now read and delete the remaining tape contents and accept on reaching the end-of-tape marker~$\lhd$.
It follows easily that $L(C)=L(A)$.
Together, the two constructions above yield the following technical result.

\begin{lemma}\label{LemNFnrNFAwtl}
Each RNFAwtl $A$ can effectively be converted into an equivalent RNFAwtl $C$
that never gets into an infinite computation
and that accepts only after reading and deleting its input completely.
In addition, if $A$ is deterministic, then so is~$C$.
\end{lemma}

Finally, we are ready to establish the following equality, which will be proved by simulation.

\begin{theorem}\label{ThmRNFAwtl}
$\mathcal{L}({\sf RNFAwtl}) = \mathcal{L}({\sf NFAwtl })$.
\end{theorem}

\noindent
{\bf Proof.} From Proposition~\ref{PropNFAinRNFA}, we know already that
$\mathcal{L}({\sf NFAwtl}) \subseteq \mathcal{L}({\sf RNFAwtl })$ holds.
Thus, it remains to prove the converse inclusion.
Accordingly, we show how to simulate an RNFAwtl by an NFAwtl.
\vspace{+2mm}

Let $A=(Q,\Sigma,\lhd,\tau,I,\delta)$ be an RNFAwtl.
By Lemma~\ref{LemNFnrNFAwtl}, we can assume that $A$ never gets into an infinite computation
and that it accepts only after reading and deleting its input completely.
We now construct an NFAwtl $B=(Q_B,\Sigma,\lhd,\tau_B,I_B,F_B,\delta_B)$ with the set of states
$Q_B = \{\,(q,\Gamma)\mid q\in Q \mbox{ and }\Gamma\subseteq\Sigma\,\}$.
The automaton $B$ uses the second component of its states
to keep track of the set of letters that may still occur on its tape.
At the beginning of its computation, $\Gamma=\Sigma$.
When $B$ simulates a step $qw\cdot\lhd \vdash_A q'w\cdot\lhd$ in which $A$ changes its state at the right sentinel because all symbols on its tape are translucent for the state~$q$,
that is, $w \in (\tau(q))^*$,
then the second component will be restricted to $\Gamma \cap \tau(q)$.
The NFAwtl $B$ is defined as follows:
\begin{itemize}
 \item $I_B = \{\,(q,\Sigma) \mid q\in I\,\}$, and
 \item $F_B = \{\,(q,\Gamma) \mid \delta(q,\lhd) = \Accept \mbox{ and }\Gamma\subseteq\Sigma\,\}$.
 \item The translucency relation $\tau_B$ is defined through
 $\tau_B((q,\Gamma)) = \tau(q)\cap\Gamma$ for all $q\in Q$ and all $\Gamma\subseteq\Sigma$, and
 \item the transition relation $\delta_B$ is initialized as follows,
 where $q\in Q$, $\Gamma\subseteq\Sigma$, and $a\in\Sigma$:
 $$
 \delta_B((q,\Gamma),a)  =  \left\{\begin{array}{rl}
                                    \{\,(q',\Gamma) \mid q'\in \delta(q,a)\,\}, & \mbox{if }a\in\Gamma \mbox{ and }\delta(q,a)\not=\emptyset,\\
                                    \emptyset,   & \mbox{if }a\in\Sigma\smallsetminus\Gamma \mbox{ or }\delta(q,a)=\emptyset.\\
                                    \end{array}\right.$$
\end{itemize}

It remains to add further transitions to $\delta_B$ that are to simulate
the transitions of the form $q'\in\delta(q,\lhd)$ of~$A$.
Assume that $q'\in\delta(q,\lhd)$.
This transition can be applied by $A$ to a configuration of the form $qw\cdot\lhd$
for which $w\in (\tau(q))^*$, and it yields the configuration $q'w\cdot\lhd$.
Thus, $A$ simply executes a change of state, but after that, it `knows'
that the word $w$ only contains occurrences of letters from the set~$\tau(q)$.
Accordingly, for each state $p\in Q$ and each letter $a\in\Sigma$,
if $q\in\delta(p,a)$, then we add the state $(q',\Gamma\cap\tau(q))$
to $\delta_B((p,\Gamma),a)$ for each subset $\Gamma$ containing the letter~$a$.
This transition allows $B$ to simulate the sequence of two transitions
$puav\cdot\lhd \vdash_A quv\cdot\lhd \vdash q'uv\cdot\lhd$,
where $u\in(\tau(p))^*$ and $u,v\in(\tau(q))^*$,
by the single transition $(p,\Gamma)uav\cdot\lhd \vdash_B (q',\Gamma\cap\tau(q))uv\cdot\lhd$.
In addition, if $q\in I$,
then the state $(q',\tau(q))$ is added to the set $I_B$,
as $A$ may start a computation by executing the step $qw\cdot\lhd \vdash_A q'w\cdot\lhd$, if $w\in(\tau(q))^*$.

It can now be checked that $B$ just simulates the computations of~$A$.
If during such a simulation, $B$ is in a state $(q,\Gamma)$
but an occurrence of a letter $c\not\in\Gamma$ is encountered,
then $B$ gets stuck and, so, rejects,
as in that situation, the letter $c$ is neither translucent for the state $(q,\Gamma)$
nor is the transition $\delta_B((q,\Gamma),c)$ defined.
It follows that $L(B)=L(A)$, which completes the proof.
\hspace*{\fill}$\Box$
\vspace{+2mm}

It remains to compare the RDFAwtl to the nr-DFAwtl, the nr-NFAwtl, and the (N)ROWJFA.
The deterministic linear language $L_2 = \{\,a^nb^n\mid n\ge 0\,\}$ is accepted by an nr-DFAwtl~\cite{otto268}.
However, it is not accepted by any NFAwtl~\cite{otto185}.
Hence, we get the following result from Lemma~\ref{LemRDFAinnrDFA}.

\begin{corollary}\label{CorRDFAinnrDFA}
$\mathcal{L}(\mbox{\sf RDFAwtl}) \subsetneq \mathcal{L}(\mbox{\sf nr-DFAwtl})$.
\end{corollary}

According to~\cite{otto270},
the language classes $\mathcal{L}(\mbox{\sf DFAwtl})$ and $\mathcal{L}(\mbox{\sf NFAwtl})$ are incomparable under inclusion
to the classes $\mathcal{L}(\mbox{\sf ROWJFA})$ and $\mathcal{L}(\mbox{\sf NROWJFA})$.
Accordingly, we also have the following incomparability result.

\begin{corollary}\label{CorRDFAvsROWJFA}
The language class $\mathcal{L}(\mbox{\sf RDFAwtl})$
is incomparable under inclusion to the classes $\mathcal{L}(\mbox{\sf ROWJFA})$ and $\mathcal{L}(\mbox{\sf NROWJFA})$.
\end{corollary}

\begin{figure}[ht]
\begin{center}
{\footnotesize
$\xymatrix@R10pt@C6pt{
 \txt{\sf CSL}\\
   \txt{\sf GCSL} \ar@{->}[u]
  & \txt{$\mathcal{L}(\mbox{\sf nr-NFAwtl})$}\ar@{->}[ul] & & \\
  \txt{\sf CFL}\ar@{->}[u]  &  & \txt{$\mathcal{L}(\mbox{\sf nr-DFAwtl})$}\ar@{->}[ul] & \txt{$\mathcal{L}(\mbox{\sf NROWJFA})$}\ar@{->}[ull]& \\
  \txt{\sf LIN}\ar@{->}[u] & \txt{$\mathcal{L}(\mbox{\sf NFAwtl}) = \mathcal{L}(\mbox{\sf RNFAwtl})$}\ar@{->}[uu] & & \\
  \txt{\sf DLIN}\ar@{->}[u] &  \txt{${\sf LRAT}$} \ar@{->}[u]
 & \txt{$\mathcal{L}(\mbox{\sf RDFAwtl})$}\ar@{->}[uu]\ar@{->}[lu] & \txt{$\mathcal{L}(\mbox{\sf ROWJFA})$}\ar@{->}[uu]\ar@{->}[uul]\\
 & & \txt{$\mathcal{L}(\mbox{\sf DFAwtl})$}\ar@{->}[u]\\
 \txt{\sf REG}\ar@{->}[uu] \ar@{->}[ruu] & \txt{$\mathcal{L}(\mbox{\sf nr-nr-NFAwtl})$}\ar@{=}[l]\ar@{=}[r] & \txt{$\mathcal{L}(\mbox{\sf nr-nr-DFAwtl})$}\ar@{->}[u] \ar@/_0.5pc/[uur]\\
}$
}
\caption{The inclusion relations between the various types of finite automata with translucent letters.}\label{FigDia2}
\end{center}
\end{figure}
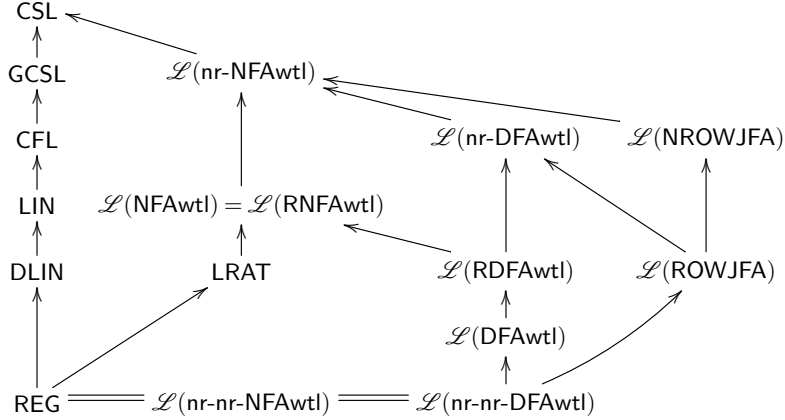

The diagram in Figure~\ref{FigDia2} summarizes the inclusion and incomparability results obtained for the various types of automata with translucent letters.
All arrows in that diagram denote proper inclusions, and classes
that are not connected by a sequence of arrows are incomparable under inclusion.
Here, ${\sf LRAT}$ denotes the class of rational trace languages (see, e.g.,~\cite{otto195}),
{\sf GCSL} is the class of growing context-sensitive languages (see, e.g.,~\cite{DaWa86}),
and {\sf (D)LIN} is the class of (deterministic) linear context-free languages.

\section{Closure Properties}\label{sec3}
In this section,
we study closure properties of the classes of languages accepted by deterministic and nondeterministic repetitive
finite automata with translucent letters.
As the class $\mathcal{L}({\sf RNFAwtl})$ of languages accepted by RNFAwtls coincides
with the class $\mathcal{L}({\sf NFAwtl})$,
based on \cite{otto195}, we obtain immediately that $\mathcal{L}({\sf RNFAwtl})$ is closed under
union, product, Kleene star, inverse projections, disjoint shuffle, and the operation of taking the commutative closure,
but it is neither closed under intersection (with regular sets), nor under complementation, nor under non-erasing morphisms.

On the other hand, for the class $\mathcal{L}({\sf DFAwtl})$, it is known~\cite{otto206}
that it is closed under complementation,
but it is not closed under any of the following operations:
union, intersection (with regular sets), product, Kleene star, reversal, alphabetic morphisms, and commutative closure.

Here, the commutative closure of a language $L$ is based on the letter-equivalence of words.
We say that two words $u$ and $v$ over an alphabet $\Sigma$ are \emph{letter-equivalent}
if, for each letter $a \in \Sigma$, $|u|_a = |v|_a$.
The \emph{commutative closure} ${\rm com}(L)$ of a language $L \subseteq \Sigma^*$
is the set of all words that are letter-equivalent to a word from~$L$, that is,
${\rm com}(L) = \{\, w \in \Sigma^* \mid \exists u \in L: u \text{ is letter-equivalent to } w\,\}.$

In addition, we are interested in the shuffle operation.
For two words $u$ and $v$ from~$\Sigma^*$,
the \emph{shuffle} $u \shuffle v$ of $u$ and $v$ is the set of words
$$u \shuffle v = \{\,u_1v_1u_2v_2\cdots u_mv_m  \mid  m\ge 1, \forall i=1,2,\ldots,m: u_i,v_i\in\Sigma^*,
   u=u_1u_2\cdots u_m, \mbox{ and } v=v_1v_2\cdots v_m\,\},$$
and the \emph{shuffle} of two languages $L_1,L_2\subseteq \Sigma^*$ is the language
 $L_1 \shuffle L_2 = \bigcup_{u\in L_1,v\in L_2} (u \shuffle v).$

We shall show that $\mathcal{L}({\sf RDFAwtl})$ is closed under complementation, left quotient with respect to a single word, and disjoint shuffle.
However, it is not closed under union, intersection (with regular sets), product, Kleene star, reversal,
alphabetic morphism, commutative closure, and shuffle.

To begin with, observe that the languages
$$L_==\{\,w\in\{a,b\}^*\mid |w|_a = |w|_b\,\} \mbox{ and }
L_{2=}=\{\,w\in\{a,b\}^*\mid 2\cdot|w|_a = |w|_b\,\}$$
are both accepted by DFAwtls.
However, their union is the language $L_\vee$ that is not even accepted by any nr-DFAwtl.
In addition, if $R$ denotes the regular language that is defined through the regular expression
$(ab)^* + (abb)^*$, then ${\rm com}(R) = L_\vee$.
Moreover, let $L_{2=}'=\{\,w\in\{c,d\}^*\mid 2\cdot|w|_c=|w|_d\,\}$.
Then, it is easily verified that the language $L_=\,\cup\,L_{2=}'$ is also accepted by a DFAwtl.
However, if $\varphi:\{a,b,c,d\}^*\to\{a,b\}^*$ denotes the alphabetic morphism that is defined through
$a\mapsto a$, $b\mapsto b$, $c\mapsto a$, and $d\mapsto b$,
then $\varphi(L_=\,\cup\,L_{2=}') = L_\vee$.
Finally, the language $L_2 = \{\,a^nb^n\mid n\ge 0\,\} = L_=\,\cap\,\left(\{a\}^*\cdot\{b\}^*\right)$
is not accepted by any NFAwtl~\cite{otto185}.
In summary, these examples yield the following non-closure properties.

\begin{corollary}\label{CorUnion}
The language class $\mathcal{L}(\mbox{\sf RDFAwtl})$ is not closed under union,
intersection (with regular sets),
alphabetic morphisms,
or commutative closure.
\end{corollary}

Next, we prove that the class $\mathcal{L}(\mbox{\sf RDFAwtl})$ is closed under the operation of complementation.

\begin{proposition}\label{PropClosCompl}
$\mathcal{L}({\sf RDFAwtl})$ is closed under complementation.
\end{proposition}

\beginproof
Let $A=(Q_A,\Sigma,\lhd,\tau_A,q_0,\delta_A)$ be an RDFAwtl.
To obtain an RDFAwtl $B=(Q_B,\Sigma,\lhd,\tau_B,q_0,\delta_B)$
accepting the complement of $L(A)$, we need to change all accepting steps into
rejecting ones and all rejecting steps into accepting ones.
For that matter, we extend the set of states of $A$ with an additional state $q_a$
that will always lead to acceptance, that is, $Q_B = Q_A \cup \{q_a\}$.
The translucency mapping $\tau_B(q)$ equals $\tau_A(q)$ for all states $q\in Q_A$,
and $\tau(q_a) = \emptyset$.
Finally, we define the transition relation of $B$ as follows,
where $q\in Q_A$ and $x\in\Sigma$:
$$\arraycolsep2pt
\begin{array}{lcl}
\delta_B(q,x) & = & \left\{ \begin{array}{ll}
                          \delta_A(q,x), & \mbox{if } \delta_A(q,x)\ne \emptyset,\\
                          q_a, &\mbox{if } \delta_A(q,x)=\emptyset \mbox{ and } x\not\in \tau(q),
                         \end{array}\right.\\[1em]
\delta_B(q,\lhd) & = & \left\{ \begin{array}{ll}
                          \delta_A(q,\lhd), & \mbox{if }\delta_A(q,\lhd)\in Q_A,\\
                          \emptyset, & \mbox{if }\delta_A(q,\lhd)=\Accept,\\
                          q_a, &\mbox{if } \delta_A(q,\lhd)=\emptyset,
                         \end{array}\right.\\[1em]
\delta_B(q_a,x)  & = & q_a,\\                            
\delta_B(q_a,\lhd) & = & \Accept.
\end{array}
$$
Now, whenever a computation of the automaton $A$ on an input word $w\in \Sigma^*$ is accepting, it is of the following form:
$$ q_0w\cdot\lhd \vdash_A^* qw'\cdot\lhd \vdash_A \Accept, \mbox{ where }q\in Q_A, w' \in (\tau(q))^*, \mbox{ and }\delta_A(q,\lhd)=\Accept.
$$
Then, the automaton $B$ will execute the computation
$ q_0w\cdot\lhd \vdash_B^* qw'\cdot\lhd \vdash_B \Reject.$
Whenever a computation of $A$ on an input word $w\in \Sigma^*$ is rejecting, then
$$ q_0w\cdot\lhd \vdash_A^* qw'\cdot\lhd \vdash_A \Reject \mbox{, where }q\in Q_A\mbox{ and } w' \in \Sigma^*.$$
Here either
\begin{enumerate}
    \item \label{case1} $w'=uxv$ for some $u \in (\tau(q))^*, v \in \Sigma^*, x\in \Sigma\smallsetminus \tau(q)$
    such that $\delta_A(q,x)=\emptyset$, or
    \item \label{case2} $w' \in (\tau(q))^*$ and $\delta_A(q,\lhd)=\emptyset$.
\end{enumerate}
In both cases, the automaton $B$ will execute an accepting computation of the form
$$ q_0w\cdot\lhd \vdash_B^* qw'\cdot\lhd \vdash_B q_aw'\cdot\lhd \vdash_B^* \Accept.$$
Thus, we see that $L(B) = \Sigma^*\smallsetminus L(A)$, the complement of the language $L(A)$.
\myendproof

The next lemma states that an RDFAwtl can be assumed to always read the very first letter during the first step of each accepting computation.
This is a purely technical result that will be useful below.

\begin{lemma}\label{LemFirstLetter}
From an RDFAwtl $A$, one can effectively construct an equivalent RDFAwtl $B$ such that,
in the first step of each computation, $B$ reads the very  first letter of the given input.
\end{lemma}

\beginproof
Let $A=(Q_A,\Sigma,\lhd,\tau_A,q_0,\delta_A)$ be an RDFAwtl, and let $L$ be the language accepted by~$A$.
By Lemma~\ref{LemNFnrNFAwtl}, we can assume that $A$ never gets into an infinite computation and that
it reads and deletes its input completely in each accepting computation.
In fact, we can even assume that $A$ reads and deletes its input completely in each and every computation,
that is, even in all its rejecting computations (see, e.g., the proof of Prop.~\ref{PropClosCompl}).
Moreover, we may assume without loss of generality that the initial state $q_0$ is not entered by any transition,
that is, this state can only occur in initial configurations of~$A$.

We now describe the RDFAwtl $B=(Q_B,\Sigma,\lhd,\tau_B,q_0,\delta_B)$ through the following definition:\hfill\\[-3mm]
$$\arraycolsep2pt
\begin{array}{clcl}
- & Q_B & = & Q_A \,\cup\, \{\,(q,a)\mid q\in Q_A \mbox{ and }a\in\Sigma \mbox{ such that }a\in\tau_A(q)\,\},\\[+2mm]
- & \tau_B(p) & = & \left\{\begin{array}{ll}
                                 \emptyset, & \mbox{if }p=q_0,\\
                                 \tau_A(p), & \mbox{if }p\in Q_A \smallsetminus\{q_0\},\\
                                 \tau_A(q),  & \mbox{if }p=(q,a) \mbox{ for some }q\in Q_A\mbox{ and }a\in\Sigma,
                             \end{array}\right.\\[+6mm]
- & \multicolumn{3}{l}{\mbox{and the transition function }\delta_B \mbox{ is defined as follows, where }q\in Q_A\smallsetminus\{q_0\}\mbox{ and }a\in\Sigma:}\\[+2mm]
  & \delta_B(q_0,a) & = & \left\{\begin{array}{ll}
                                  \delta_A(q_0,a), & \mbox{if }a\not\in\tau_A(q_0),\\
                                  (q_0,a), & \mbox{if }a\in\tau_A(q_0),
                                 \end{array}\right.\\[+4mm]
  & \delta_B(q_0,\lhd) & = & \delta_A(q_0,\lhd),\\[+2mm]
  & \delta_B(q,b) & = & \delta_A(q,b) \mbox{ for all }q\in Q_A\smallsetminus\{q_0\}\mbox{ and }b\in\Sigma\,\cup\,\{\lhd\},\\[+2mm]
  & \delta_B((q,a),b) & = & \left\{\begin{array}{ll}
                        (\delta_A(q,b),a),  & \mbox{if }b\not=a \mbox{ and }a\in\tau_A(\delta_A(q,b)),\\
                        \delta_A(\delta_A(q,b),a), & \mbox{if }b\not=a \mbox{ and }a\not\in\tau_A(\delta_A(q,b)),\\
                        \emptyset,                 & \mbox{if }b=a,
                        \end{array}\right.\\[+6mm]
  & \delta_B((q,a),\lhd) & = & \left\{\begin{array}{ll}
                (\delta_A(q,\lhd),a),  & \mbox{if }a\in\tau_A(\delta_A(q,\lhd)),\\
                \delta_A(\delta_A(q,\lhd),a), & \mbox{if }a\not\in\tau_A(\delta_A(q,\lhd)).\\
                        \end{array}\right.
\end{array}$$
The states of the form $(q,a)$ are used to encode the fact that $A$ is in state~$q$ and that the first letter
on the tape was an~$a$,
which, however, has not yet been read by~$A$ (but was already read by $B$).
Hence, by our above assumptions on the computations of~$A$,
we see that, if $B$ is in state $(q,a)$ reading the sentinel~$\lhd$,
then $\delta_A(q,\lhd)\in Q$ holds.

From the above definition, it follows immediately that, in each computation,
the RDFAwtl~$B$ reads and deletes the first letter on the tape.
Moreover, the initial state $q_0$, which is not entered by any transition of~$A$,
is not entered by any transition of~$B$, either.
Now, by comparing the computations of $B$ on a given word $w$ to that of $A$ on $w$,
it can be verified that $B$ is equivalent to~$A$,
that is, that $L(B) = L(A)$ holds.
\myendproof

For a language $L\subseteq \Sigma^*$ and a word $w\in\Sigma^*$,
the \emph{left quotient} of $L$ with respect to $w$ is the language
$$w\leftthreetimes L = \{\,z\in\Sigma^* \mid wz\in L\,\}.$$
If $L$ is accepted by an RDFAwtl $A=(Q,\Sigma,\lhd,\tau,q_0,\delta)$, then by Lemma~\ref{LemFirstLetter}, we can assume
that $A$ always reads the first letter of the given input during the first step of each computation.
Hence, for each letter $a\in\Sigma$, the RDFAwtl $B_a = (Q,\Sigma,\lhd,\tau,\delta(q_0,a),\delta)$
accepts the language $a\leftthreetimes L$,
which shows that $a\leftthreetimes L$ is accepted by an RDFAwtl.
By induction on $|w|$, it now follows that $w\leftthreetimes L$ is accepted by an RDFAwtl
for each word~$w$.

\begin{corollary}\label{CorClosureLeftQuotient}
The language class $\mathcal{L}(\mbox{\sf RDFAwtl})$ is closed under the operation of
taking the left quotient with respect to a single word.
\end{corollary}

To derive the other non-closure properties stated above, we need the following technical results.

\begin{lemma}\label{LemLc}
None of the following languages is accepted by an RDFAwtl:\hfill\\[+1mm]
$\begin{array}{clcl}
{\rm (a)} & L_c &  =  & \{\,wc\mid w\in\{a,b\}^*, |w|_a\ge |w|_b\,\},\\
{\rm (b)} & \multicolumn{3}{l}{(L_c^R)^+ \mbox{ and } (L_c^R)^*, \mbox{ and }} \\
{\rm (c)} & L & = & \{\,w\in\{a,b\}^* \mid |w|_a \le |w|_b \le 2\cdot|w|_a\,\}.\\
\end{array}$
\end{lemma}

\beginproof
(a)
Let $\Sigma = \{a,b,c\}$.
For deriving a contradiction, we assume that
$A=(Q,\Sigma,\lhd,\tau,q_0,\delta)$ is an RDFAwtl such that $L(A) = L_c$.
Without loss of generality, we may assume that $A$ reads and deletes its input completely during each accepting computation.

For each $n\ge 1$, the word $a^nb^nc$ belongs to the language~$L_c$.
Accordingly, the computation of $A$ on input $w_n = a^nb^nc$ is of the form
$q_0a^nb^nc\cdot\lhd \vdash_A^* q_f\cdot\lhd \vdash_A \Accept,$
where $q_f\in Q$.
Thus, in particular,
the single occurrence of the letter $c$ is read and deleted during this computation.
Accordingly, this computation can be written as follows:
$$q_0a^nb^nc\cdot\lhd \vdash_A^{*} qa^{n-i}b^{n-j}c\cdot\lhd \vdash_A q'a^{n-i}b^{n-j}\cdot\lhd \vdash_A^* q_f\cdot\lhd \vdash_A \Accept,$$
where $q,q'\in Q$, $0\le i,j\le n$, and $\delta(q,c)=q'$.
Then $A$ can also execute the following computation:
$$q_0a^ib^jca^{n-i}b^{n-j}\cdot\lhd \vdash_A^{*} qca^{n-i}b^{n-j}\cdot\lhd \vdash_A q'a^{n-i}b^{n-j}\cdot\lhd \vdash_A^*q_f\cdot\lhd \vdash_A \Accept.$$
However, the word $a^ib^jca^{n-i}b^{n-j}$ is not an element of the language $L_c$ unless $i=n$ and $j=n$,
which means that, in the accepting computation above, $A$ first reads and deletes all occurrences of the letters $a$ and $b$
before it reads the single occurrence of the letter~$c$.
In particular, it follows that this computation has the form
$q_0a^nb^nc\cdot\lhd \vdash_A^*qc\cdot\lhd \vdash_A q'\cdot\lhd \vdash_A^*q_f\cdot\lhd \vdash_A \Accept.$
\vspace{+2mm}

Now, let $n>|Q|$.
If the computation $q_0a^nb^nc\cdot\lhd \vdash_A^* qc\cdot\lhd$ begins
with $|Q|$ many steps that each read an occurrence of the letter~$a$,
then there is a state $q\in Q$ that is used twice during these steps,
that is
$$q_0a^nb^nc\cdot\lhd \vdash_A^* qa^{n-i}b^nc\cdot\lhd \vdash_A^+ qa^{n-i-j}b^nc\cdot\lhd \vdash_A^* \Accept,$$
where $i\ge 0$, $j\ge 1$, and $i+j\le |Q|$.
But then $A$ can also execute the following accepting computation:
$$q_0a^{n-j}b^nc\cdot\lhd \vdash_A^*qa^{n-i-j}b^nc\cdot\lhd \vdash_A^* \Accept.$$
However, as $j\ge 1$, the word $a^{n-j}b^nc$ does not belong to the language~$L_c$, a contradiction.
This implies that, within the first $|Q|$ many steps in the accepting computation considered,
an occurrence of the letter~$b$ is read and deleted.
\vspace{+2mm}

Let us now consider the first step in the above computation
during which an occurrence of the letter~$b$ is deleted:
$$q_0a^nb^nc\cdot\lhd \vdash_A^* q_1a^{n-i}b^nc\cdot\lhd \vdash_A q_2a^{n-i}b^{n-1}c\cdot\lhd,$$
where $0\le i< |Q|$, $q_1,q_2\in Q$, $a\in\tau(q_1)$, and $\delta(q_1,b)=q_2$.
Given the input $a^nc\in L_c$,
$A$ will also accept.
However, as $A$ is deterministic, the corresponding accepting computation begins with 
$$q_0a^nc\cdot\lhd \vdash_A^* q_1a^{n-i}c\cdot\lhd,$$
where $a\in\tau(q_1)$.

If $c\in\tau(q_1)$, then
together with the partial computation $q_0a^nbc\cdot\lhd \vdash_A^* q_1a^{n-i}bc\cdot\lhd \vdash_A q_2a^{n-i}c\cdot\lhd,$
the automaton $A$ could also execute the following partial computation:
$$q_0a^ncb\cdot\lhd \vdash_A^* q_1a^{n-i}cb\cdot\lhd \vdash_A q_2a^{n-i}c\cdot\lhd.$$
As $a^nbc\in L_c$, the former computation leads to acceptance and, hence, so does the latter.
This yields a contradiction, as $a^ncb\not\in L_c$.
Hence, it follows that $c\not\in\tau(q_1)$.

This implies that, in the above configuration $q_1a^{n-i}c\cdot\lhd$, $A$ reads and deletes the letter~$c$,
that is, $\delta(q_1,c) = q'$ for some $q'\in Q$.
Thus, we obtain the accepting computation
$$q_0a^nc\cdot\lhd \vdash_A^* q_1a^{n-i}c\cdot\lhd \vdash_A q'a^{n-i}\cdot\lhd\vdash_A^* \Accept.$$
But then, we also obtain the accepting computation
$$q_0a^ica^{n-i}\cdot\lhd \vdash_A^* q_1ca^{n-i}\cdot\lhd \vdash_A q'a^{n-i}\cdot\lhd \vdash_A^* \Accept,$$
which yields a contradiction, as $a^ica^{n-i}\not\in L_c$.
In summary, this shows that the language $L_c$ is not accepted by any RDFAwtl.
\vspace{+2mm}

\noindent
(b) Assume to the contrary that $(L_c^R)^+$ or $(L_c^R)^*$ is accepted by an RDFAwtl.
Lemma~\ref{LemFirstLetter} implies that the language
$$\begin{array}{lclcl}
L & = & c\leftthreetimes (L_c^R)^+  & = & c\leftthreetimes (L_c^R)^*\\
  & = &\multicolumn{3}{l}{\{\,w_1cw_2c\cdots cw_k\mid k\ge 1 \mbox{ and }
  \forall i=1,2,\ldots,k: w_i\in\{a,b\}^* \wedge |w_i|_a \ge |w_i|_b\,\} }
\end{array}$$
is also accepted by an RDFAwtl $A=(Q,\Sigma,\lhd,\tau,q_0,\delta)$, where $\Sigma=\{a,b,c\}$.
For all $m\ge 0$, the word $w(m) = a^mb^mcab$ belongs to the language~$L$, which means
that the computation of $A$ on input $w(m)$ is accepting.
However, using pumping arguments as in the proof of (a),
it can now be shown that, together with the words $w(m)$,
the RDFAwtl $A$ also accepts some words that do not belong to the language $(L_c^R)^*$.
Accordingly, the languages $(L_c^R)^+$ and $(L_c^R)^*$ are not accepted by any RDFAwtls.
\vspace{+2mm}

\noindent
(c) By using pumping arguments, it can be proved that an RDFAwtl accepting the language 
$$L  =  \{\,w\in\{a,b\}^* \mid |w|_a \le |w|_b \le 2\cdot|w|_a\,\}$$
also accepts some words that do not belong to this language.
Again, this shows that this language is not accepted by any RDFAwtl.
\myendproof

It is easily seen that the languages
$$L_c^R = \{\,cw \mid w\in\{a,b\}^*, |w|_a\ge |w|_b\,\},\,
L_\ge = \{\,w\in\{a,b\}^*\mid |w|_a\ge |w|_b\,\}, \mbox{ and } \{c\}$$
are all accepted by DFAwtls.
On the other hand, 
$L = L_= \shuffle L_{2=} = \{\,w\in\{a,b\}^* \mid |w|_a \le |w|_b \le 2\cdot|w|_a\,\}$
is not.
Hence, Lemma~\ref{LemLc} shows the following.

\begin{corollary}\label{CorProduct}
The language class $\mathcal{L}(\mbox{\sf RDFAwtl})$ is not closed under reversal, (disjoint) product, Kleene plus,
Kleene star,
or shuffle.
\end{corollary}

Finally, the class $\mathcal{L}(\mbox{\sf RDFAwtl})$ is closed under a restricted variant of the shuffle operation.
If $\Sigma$ is an alphabet and $\Delta$ is a subalphabet of~$\Sigma$,
we shall use $P_\Delta$ to denote the
projection $P_{\Delta}: \Sigma \to \Delta$ that maps each letter from~$\Delta$ to itself
and each letter from $\Sigma\smallsetminus \Delta$
to the empty word~$\lambda$.
The projection $P_\Delta$ can be extended to words and languages in a natural way.

If a word $w \in (\Sigma_A \cup \Sigma_B)^*$ is in the set $u \shuffle v$ for some words $u\in \Sigma_A^*$ and $v \in \Sigma_B^*$,
where $\Sigma_A$ and $\Sigma_B$ are two disjoint alphabets,
then $P_{\Sigma_A}(w) = u$ and $P_{\Sigma_B}(w) = v$.
The shuffle of two words or languages over disjoint alphabets is called a \emph{disjoint shuffle}.

\begin{proposition}\label{PropDisjShuffle}
The language class $\mathcal{L}(\mbox{\sf RDFAwtl})$ is closed under disjoint shuffle.
\end{proposition}

\beginproof
Let $\Sigma_A$ and $\Sigma_B$ be two disjoint alphabets,
let $A = (Q_A,\Sigma_A,\lhd,\tau_A,q_0^{(A)},\delta_A)$ be an RDFAwtl on $\Sigma_A$ that accepts a language $L(A) = L_A$,
and let $B=(Q_B,\Sigma_B,\lhd,\tau_B,q_0^{(B)},\delta_B)$ be an RDFAwtl on $\Sigma_B$ that accepts a language $L(B)=L_B$.
We shall construct an RDFAwtl $M$ for the disjoint shuffle $L = L_A \shuffle L_B$ of $L_A$ and~$L_B$.

By Lemma~\ref{LemNFnrNFAwtl}, we can assume without loss of generality that $A$ never gets into an infinite computation
and that it reads and deletes its input completely in each accepting computation.
The RDFAwtl $M=(Q,\Sigma,\lhd,\tau,q_0,\delta)$ is constructed as follows, where we assume
that the sets of states $Q_A$ and $Q_B$ are disjoint:\hfill\\[+2mm]
$\arraycolsep2pt
\begin{array}{clcll}
- & Q & = & \multicolumn{2}{l}{Q_A \,\cup\,Q_B\mbox{ and }q_0 = q_0^{(A)},}\\[+2mm]
- & \tau(q) & = & \left\{\begin{array}{ll}
                         \tau_A(q) \cup \Sigma_B, & \mbox{if }q\in Q_A,\\
                         \tau_B(q),               & \mbox{if }q\in Q_B,\\
                        \end{array}\right.\\[+4mm]
- & \delta(q,a) & = & \left\{\begin{array}{ll}
                          \delta_A(q,a), & \mbox{if }q\in Q_A\mbox{ and }a\in \Sigma_A,\\
                          \delta_B(q,a), & \mbox{if }q\in Q_B\mbox{ and }a\in \Sigma_B,\\
                          \emptyset,     & \mbox{otherwise},\\
                           \end{array}\right.\\[5mm]
 & \delta(q,\lhd) & = & \left\{\begin{array}{ll}
                           \delta_A(q,\lhd), & \mbox{if }q\in Q_A\mbox{ and }\delta_A(q,\lhd)\not=\Accept,\\
                           q_0^{(B)},        & \mbox{if }q\in Q_A\mbox{ and }\delta_A(q,\lhd) = \Accept,\\
                           \delta_B(q,\lhd), & \mbox{if }q\in Q_B.
                           \end{array}\right.\\
\end{array}$\hfill\\

For an input $w \in u\shuffle v$, where $u\in\Sigma_A^*$ and $v\in\Sigma_B^*$,
the RDFAwtl $M$ first simulates the computation of~$A$ on~$u$, ignoring all occurrences of letters from~$\Sigma_B$.
When~$A$ accepts, then $M$ simulates the computation of~$B$ on~$v$.
As~$A$ reads and deletes all letters of $u$ in its accepting computation,
it is obvious that $M$ accepts on input $w$ if and only if $A$ accepts on input $u$ and $B$ accepts on input~$v$.
It follows that $L(M) =L(A) \shuffle L(B) = L_A \shuffle L_B$.
Hence, $\mathcal{L}(\mbox{\sf RDFAwtl})$ is closed under the operation of disjoint shuffle.
\myendproof

\section{Conclusion}\label{sec4}
Concerning the complexity of the membership problem, it is easily seen that the algorithm
for the membership problem of a DFAwtl presented by
Nagy and Kov{{\'a}}cs in~\cite{NagKov14} applies to an RDFAwtl as well.
This yields the following result. Note that the complexity of the membership problem of a DFAwtl is measured using a logarithmic cost of instructions.

\begin{corollary}\label{CorMembership}
The membership problem for an RDFAwtl is
decidable in time ${O}(n\cdot \log n)$.
\end{corollary}

Furthermore, emptiness and finiteness are decidable for RDFAwtls, as they are decidable for NFAwtls~\cite{otto206}.
As $\mathcal{L}(\mbox{\sf RDFAwtl})$ is effectively closed under complementation,
universality is also decidable for these automata.
On the other hand,
the problem of deciding whether the language accepted by a given RDFAwtl has a non-empty intersection
with a given regular language is undecidable,
as this problem is already undecidable for DFAwtls.
The same holds for the problem of deciding whether the language accepted by a given RDFAwtl
contains (or is contained in) a given regular language,
and therewith, the inclusion problem for RDFAwtls is undecidable, too.
However, it remains open whether the equivalence problem is decidable for RDFAwtls.

In summary, our results show that the RDFAwtl is just slightly more expressive than the DFAwtl,
but it seems to have the same closure and non-closure properties,
and the same decidability and undecidability results seem to hold.
Moreover, we have seen that, when we add the property of being non-returning to the DFAwtl (or the NFAwtl),
we actually weaken the model.
When we add the property of repetitiveness to the DFAwtl,
then we obtain a language class that is just a bit more expressive than the DFAwtl,
while in the nondeterministic case, this generalization has no effect on the expressive capacity of the model.
However, when we add both these properties, repetitiveness and the property of being non-returning,
then the resulting types of automata, that is, the nr-DFAwtl and the nr-NFAwtl, have indeed a much larger expressive capacity
than the original models.
Thus, it is really the combination of these two properties that implicates the enormous increase in the expressive capability
from the DFAwtl and the NFAwtl to the nr-DFAwtl and the nr-NFAwtl.

\bibliographystyle{eptcs.bst}
\bibliography{NCMA2024MO}

\end{document}